\documentclass[12pt]{iopart}

\expandafter\let\csname equation*\endcsname\relax
\expandafter\let\csname endequation*\endcsname\relax
\usepackage{amsmath,amssymb,bm}
\usepackage{graphicx,hyperref,color}

\usepackage{iopams}  

\begin{document}

\title[Demonstration of a dual-pass differential Fabry--Perot interferometer]{Demonstration of a dual-pass differential Fabry--Perot interferometer for future interferometric space gravitational wave antennas}

\author{Koji Nagano\textsuperscript{1,2}, Hiroki Takeda\textsuperscript{3}, Yuta Michimura\textsuperscript{3}, Takashi Uchiyama\textsuperscript{4} \& Masaki Ando\textsuperscript{3,5}}

\address{$^1$ KAGRA Observatory, Institute for Cosmic Ray Research, The University of Tokyo, Kashiwa, Chiba 277-8582, Japan}
\address{$^2$ Institute of Space and Astronautical Science, Japan Aerospace Exploration Agency, Sagamihara, Kanagawa 252-5210, Japan}
\address{$^3$ Department of Physics, Graduate School of Science, The University of Tokyo, Bunkyo, Tokyo 113-0033, Japan}
\address{$^4$ KAGRA Observatory, Institute for Cosmic Ray Research, The University of Tokyo, Hida, Gifu 506-1205, Japan}
\address{$^5$ Research Center for the Early Universe, The University of Tokyo, Bunkyo, Tokyo 113-0033, Japan}

\ead{knagano@ac.jaxa.jp}
\vspace{10pt}
\begin{indented}
\item[]August 2020
\end{indented}

\begin{abstract} 
A dual-pass differential Fabry--Perot interferometer (DPDFPI) is one candidate of the interferometer configurations utilized in future Fabry--Perot type space gravitational wave antennas, such as Deci-hertz Interferometer Gravitational Wave Observatory. In this paper, the working principle of the DPDFPI has been investigated and necessity to adjust the absolute length of the cavity for the operation of the DPDFPI has been found. In addition, using the 55-cm-long prototype, the operation of the DPDFPI has been demonstrated for the first time and it has been confirmed that the adjustment of the absolute arm length reduces the cavity detuning as expected. This work provides the proof of concept of the DPDFPI for application to the future Fabry--Perot type space gravitational wave antennas. 
\end{abstract}

%
%
%
%
%

\section{Introduction}

The first detection of the gravitational wave from the black hole binary by the Advanced Laser Interferometer Gravitational-Wave Observatory (aLIGO) opened the era of the gravitational wave physics and astronomy \cite{Abbott2016}. The first detection has been followed by many detections of the gravitational wave from the black hole and neutron star binaries \cite{Abbott2019a, Abbott2017a}. The gravitational wave detections and its electromagnetic followup observations have already provided significant physical and astronomical information \cite{Abbott2017b,Abbott2017c}. 

For further expansion of the gravitational wave physics and astronomy, we need to observe manifold classes of the gravitational wave objects. In other words, expanding the observation frequency is required \cite{Sesana2016, Vitale2016, Isoyama2018}. Although stellar mass objects, $\sim$1-$100$ $M_{\odot}$ ($M_{\odot}$ is the solar mass), have been observed by aLIGO around 100 Hz, a relatively heavy object, e.g. $\sim$$10^3$ $M_{\odot}$ is still attractive observational target around 0.1 Hz \cite{Matsubayashi2004,Reisswig2013}.

The upper observable mass bound, i.e. lower limit of the observation frequency range, of aLIGO and the other ground based detector is mainly limited by ground motion \cite{Hughes1998,Abbott2004}. Thus, in order to observe the heavy objects, space gravitational wave antennas have been proposed, such as Laser Interferometer Space Antenna \cite{Amaro-Seoane2017}, Big Bang Observer \cite{Harry2006}, TianQin \cite{Luo2016}, Taiji \cite{Hu2017}, TianGO \cite{Kuns2019}, and DECi-hertz Interferometer Gravitational Wave Observatory (DECIGO) \cite{Kawamura2011}. 

Among them, DECIGO and its precursor proposal, B-DECIGO \cite{Kawamura2020}, are planning to utilize a Fabry--Perot interferometer to enhance the sensitivity around 0.1 Hz. The Fabry--Perot interferometer gives DECIGO the possibility even to observe a stochastic gravitational wave background generated in the early Universe \cite{Smith2006,Garcia-Bellido2008,Kuroyanagi2011,Alabidi2012}. Therefore, when the Fabry--Perot type space gravitational wave observatory is realized, new gravitational wave physical and astronomical knowledge will be provided by the observation of the astronomical objects that has not been detected.

There are, however, some challenges for use of the Fabry--Perot cavity in space mission. One of the challenges is how to ensure the redundancy with as a small number of components as possible. In space missions, the total mass of the components is critical since it is strongly restricted by the ability of the launch system. One proposed solution to ensure the redundancy is a dual-pass Fabry--Perot interferometer configuration \cite{Kawamura2011}. In the configuration, the Fabry--Perot cavity is designed to be critically coupled and two lasers from two sources are injected to the cavity from both cavity mirrors. Consequently, the cavity signal can be measured with both lasers. As a result, the redundancy of the interferometer operation is provided by the minimum number of the mirrors.

\begin{figure}[htb]
  \centering
  \includegraphics[width=100mm]{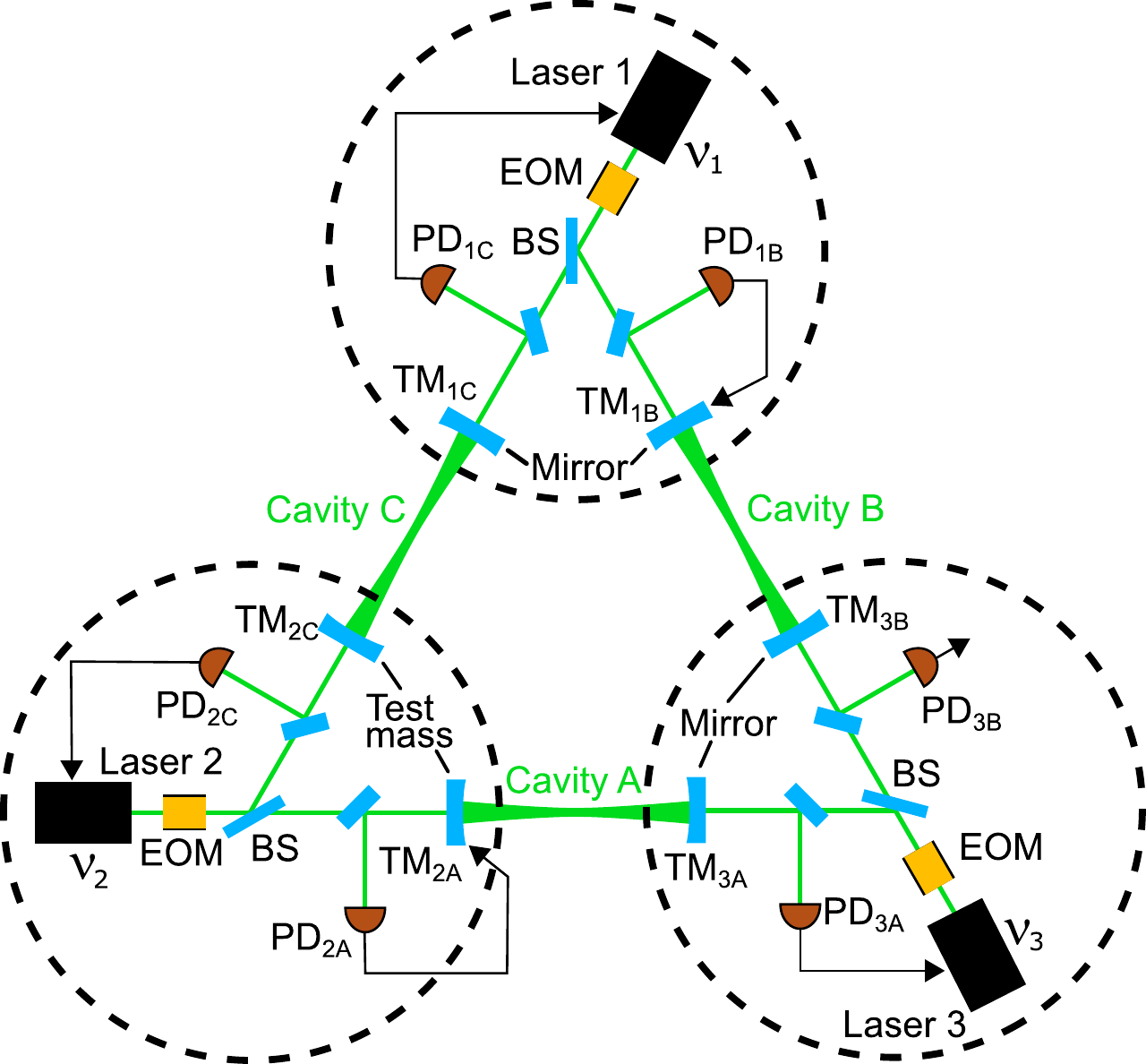}
  \caption{Schematic view of the DPDFPI considered for DECIGO. PD is a photodetector, EOM is an electro-optic modulator, BS is a beam splitter, and TM is a test mass. Instruments in the dashed circle are placed in one station, i.e. one satellite in the space detector. For decoupling, the frequency of the lasers are shifted for each other and the polarization of the lasers input to one cavity is orthogonalised.}
  \label{fig:DPDFPI}
\end{figure}

For further proceedings of the dual-pass Fabry--Perot interferometer concept in the gravitational wave detector, it was necessary to investigate a realistic proposal on the concept. In this paper, we propose a realistic dual-pass Fabry--Perot interferometer interferometer configuration, a \textit{dual-pass differential Fabry--Perot Interferometer} (DPDFPI). Figure~\ref{fig:DPDFPI} shows the schematic of the DPDFPI. In the DPDFPI, the gravitational wave signal is mainly obtained with a differential Fabry--Perot interferometer, which is adopted in some ground-based gravitational wave detectors \cite{Abramovici1996, Ohashi2003}. As another configuration with the dual-pass Fabry--Perot interferometer, a back-linked Fabry--Perot interferometer was also proposed \cite{Izumi2020}. It uses two laser sources in one satellite for each cavity metrology and the gravitational wave signal is obtained by making two lasers interfere in one satellite. Compared with the back-linked Fabry--Perot interferometer, the DPDFPI requires the relatively simple optical configuration without the back-link interferometer. However, in the DPDFPI shown in fig.~\ref{fig:DPDFPI}, we need to consider a new control topology among three cavities peculiar to the DPDFPI since the cavity mirrors are shared with each other interferometer. In this paper, we analytically investigate the working principle of the DPDFPI for the first time and show the requirement of the cavity length adjustment for the operation of the DPDFPI. Moreover, we constructed the first experimental prototype of the DPDFPI for its proof of concept.

%
%


\section{Formalization of the dual-pass differential Fabry--Perot interferometer} \label{ss:DPDFPI}

\begin{table}
\caption{Signal flow of the DPDFPI shown in fig. \ref{fig:DPDFPI}. The signal obtained with the photodetectors in the first column corresponds to the sensing objects in the second column and is fed back to the controller in the third column.}
\label{tab:signal}
\centering
\begin{tabular}{lll}
\\
\br
Photodetector & Sensing object & Controller \\
\mr
PD$_\mathrm{1B}$ & Laser 1, Cavity B & TM$_\mathrm{1B}$ \\
PD$_\mathrm{1C}$ & Laser 1, Cavity C & Laser 1 \\
PD$_\mathrm{2C}$ & Laser 2, Cavity C & Laser 2 \\
PD$_\mathrm{2A}$ & Laser 2, Cavity A & TM$_\mathrm{2A}$ \\
PD$_\mathrm{3A}$ & Laser 3, Cavity A & Laser 3 \\
PD$_\mathrm{3B}$ & Laser 3, Cavity B & --- \\
\br
\end{tabular}
\end{table}

\begin{figure}[hb]
  \centering
  \includegraphics[width=100mm]{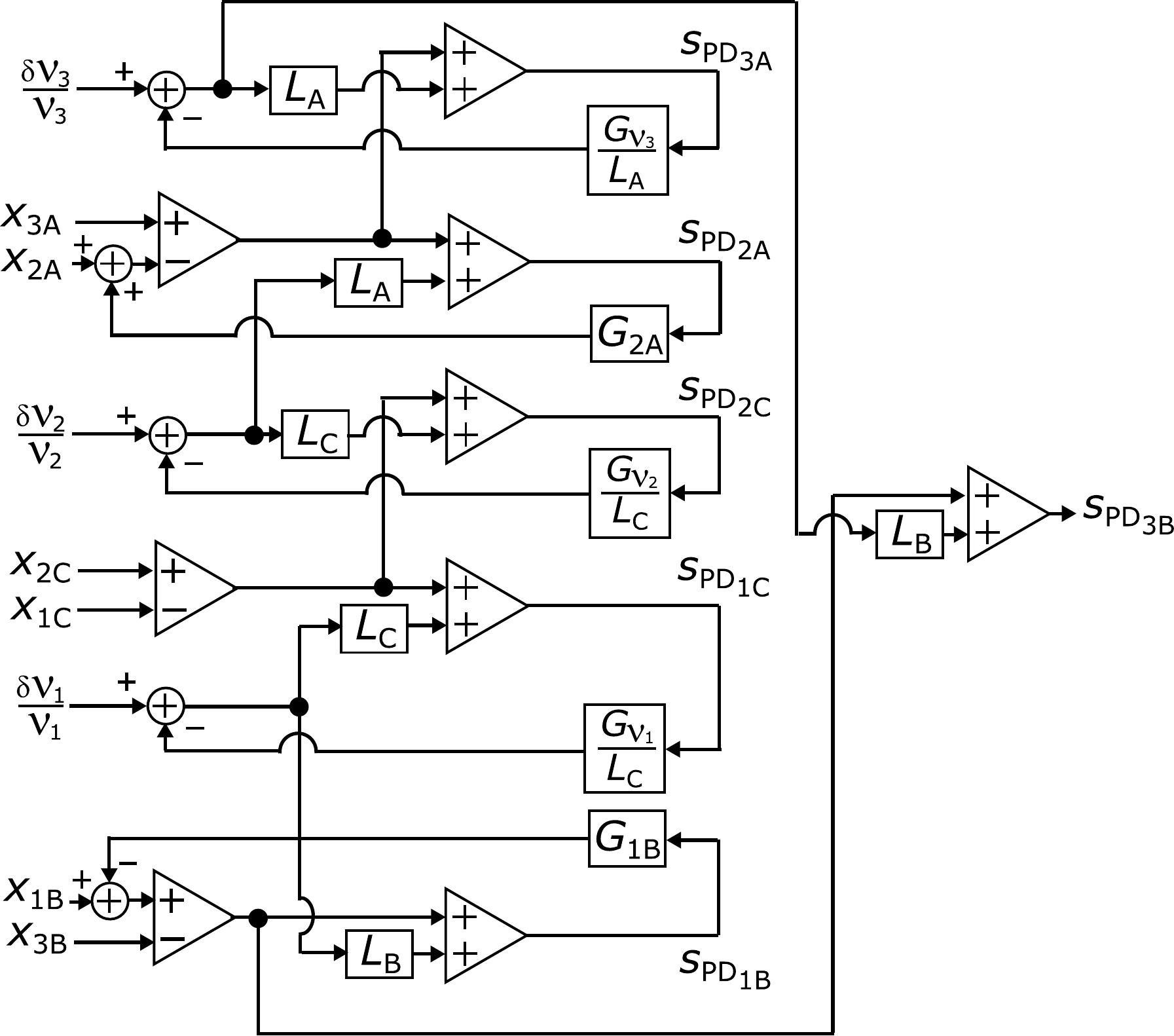}
  \caption{Block diagram of the DPDFPI shown in fig. \ref{fig:DPDFPI}. $x_{i\alpha}$ ($i = 1,2,3$ and $\alpha = \mathrm{A},\mathrm{B},\mathrm{C}$) are the longitudinal displacement of each test mass, $\nu_i$ are the laser frequency of the Laser $i$, $\delta \nu_i$ are its fluctuation, $L_\mathrm{\alpha}$ are the cavity length of the Cavity $\alpha$, $s_{\mathrm{PD}_{i\alpha}}$ are the signal obtained with each photodetector, and $G$ are the open loop gain.}
  \label{fig:BD1}
\end{figure}

We present the working principle of the DPDFPI using the block diagram. Figure \ref{fig:BD1} shows the block diagram of the DPDFPI shown in fig. \ref{fig:DPDFPI}. For the measurement with the Fabry--Perot cavity, we need to make it resonate using, for example, Pound--Drever--Hall technique \cite{Drever1983}. Usually, servo system is used to keep the resonance. Notice that the frequency of the lasers in the DPDFPI are shifted for each other for decoupling. The signal flow of the DPDFPI shown in fig. \ref{fig:DPDFPI} is presented in table \ref{tab:signal}. The length signal, i.e. the resonant frequency, of the Cavity C is fed back to the Laser 1 and 2. With this feed back system, the frequency of the Laser 1 and 2 are controlled at the resonant frequency of the Cavity C. Then, the resonant frequency of the Cavity A and B are compared with the frequency of the Laser 1 and 2, respectively, and are controlled by actuating the position of the mirrors, TM$_\mathrm{2A}$ and TM$_\mathrm{1B}$, respectively. In addition, the length signal of the Cavity A is fed back to the Laser 3 with the consequence that the frequency of the Laser 3 is controlled at the resonant frequency of the Cavity A. Thanks to the above feedback system, the Laser 1 (2) resonates with the Cavity B and C (A and C) and the Laser 3 resonates with the Cavity A. One consideration is that the length signal measured with the PD$_\mathrm{3A}$ cannot be fed back to the length of the Cavity B or the frequency of the Laser 2 since the feedback paths are occupied by the other signals. Therefore, we need some method to make the Laser 3 resonate with the Cavity B as discussed later. If the Laser 3 resonates with the Cavity B, the obtained signals in the feedback system shown in fig. \ref{fig:BD1} are expressed as,
\begin{align}
s_{\mathrm{PD_{1B}}} = & \frac{1}{1 + G_\mathrm{1B}} 
	\left( 
		- \frac{G_{\nu_1}}{1+G_{\nu_1}} \frac{L_\mathrm{B}}{L_\mathrm{C}}\Delta x_\mathrm{C} 
		+ \Delta x_\mathrm{B}
		+ \frac{1}{1 + G_{\nu_1}} L_\mathrm{B}\frac{\delta \nu_1}{\nu_1} 
	\right) ,\\
s_{\mathrm{PD_{1C}}} = & \frac{1}{1+G_{\nu_1}}
	\left(
		\Delta x_\mathrm{C} 
		+ L_\mathrm{C}\frac{\delta \nu_1}{\nu_1} 
	\right), \\
s_{\mathrm{PD_{2C}}} = & \frac{1}{1 + G_{\nu_2}}
	\left(
		\Delta x_\mathrm{C} 
		+  L_\mathrm{C}\frac{\delta \nu_2}{\nu_2} 
	\right), \\
s_{\mathrm{PD_{2A}}} = & \frac{1}{1 + G_\mathrm{2A}}
	\left(
		- \frac{G_{\nu_2}}{1+G_{\nu_2}}\frac{L_\mathrm{A}}{L_\mathrm{C}} \Delta x_\mathrm{C} 
		+ \Delta x_\mathrm{A} 
		+ \frac{1}{1+G_{\nu_2}} L_\mathrm{A}\frac{\delta \nu_2}{\nu_2} 
	\right),  \\
s_{\mathrm{PD_{3A}}} = & \frac{1}{1 + G_{\nu_3}}
	\left(
		\frac{G_{\nu_2}}{1+G_{\nu_2}}\frac{G_\mathrm{2A}}{1+G_\mathrm{2A}} \frac{L_\mathrm{A}}{L_\mathrm{C}}\Delta x_\mathrm{C}
		+ \frac{1}{1+G_\mathrm{2A}} \Delta x_\mathrm{A} \right. \nonumber \\
		& \hspace{2cm} \left.+ L_\mathrm{A}\frac{\delta \nu_3}{\nu_3} 
		-\frac{1}{1+G_{\nu_2}} \frac{G_\mathrm{2A}}{1+G_\mathrm{2A}} L_\mathrm{A}\frac{\delta \nu_2}{\nu_2} 
	\right) \label{eq:s_PD_3A}, \\
s_{\mathrm{PD_{3B}}} = & 
		\left(- \frac{G_{\nu_2}}{1 + G_{\nu_2}}\frac{G_\mathrm{2A}}{1 + G_\mathrm{2A}}\frac{G_{\nu_3}}{1 + G_{\nu_3}}
			+ \frac{G_{\nu_2}}{1 + G_{\nu_2}}\frac{G_\mathrm{1B}}{1 + G_\mathrm{1B}} \right) \frac{L_\mathrm{B}}{L_\mathrm{C}}\Delta x_\mathrm{C} \nonumber \\
		& \ - \frac{1}{1 + G_\mathrm{2A}}\frac{G_{\nu_3}}{1 + G_{\nu_3}} \frac{L_\mathrm{B}}{L_\mathrm{A}}\Delta x_\mathrm{A} 
		+ \frac{1}{1 + G_\mathrm{1B}} \Delta x_\mathrm{B}
		+ \frac{1}{1 + G_{\nu_3}} L_\mathrm{B}\frac{\delta \nu_3}{\nu_3} \nonumber \\
		& \ - \frac{1}{1 + G_{\nu_1}}\frac{G_\mathrm{1B}}{1 + G_\mathrm{1B}} L_\mathrm{B}\frac{\delta \nu_1}{\nu_1} 
		+ \frac{1}{1 + G_{\nu_2}}\frac{G_\mathrm{2A}}{1 + G_\mathrm{2A}}\frac{G_{\nu_3}}{1 + G_{\nu_3}} L_\mathrm{B}\frac{\delta \nu_2}{\nu_2}, \label{eq:s_PD_3B}
\end{align}
where $\Delta x_\mathrm{A} \equiv x_\mathrm{3A}-x_\mathrm{2A}$, $\Delta x_\mathrm{B} \equiv x_\mathrm{1B}-x_\mathrm{3B}$, and $\Delta x_\mathrm{C} \equiv x_\mathrm{2C}-x_\mathrm{1C}$ are the cavity length fluctuation, $x_{i\alpha}$ ($i = 1,2,3$ and $\alpha = \mathrm{A},\mathrm{B},\mathrm{C}$) are the longitudinal displacement of each test mass, $\nu_i$ are the laser frequency of the Laser $i$, $\delta \nu_i$ are its fluctuation, $L_\mathrm{\alpha}$ are the cavity length of the Cavity $\alpha$, $s_{\mathrm{PD}_{i\alpha}}$ are the signal obtained with each photodetector, and $G$ are the open loop gain. 
When $|G_{\nu_i}| \gg 1$ and $|G_{i\alpha}| \ll 1$ (or $G_\mathrm{2A} \simeq G_\mathrm{1B}$), $s_{\mathrm{PD_{2A}}}$, $s_{\mathrm{PD_{1B}}}$, and $s_{\mathrm{PD_{3B}}}$, are denoted as
\begin{align}
s_{\mathrm{PD_{1B}}} = & \frac{1}{1 + G_\mathrm{1B}} 
	\left( 
		- \Delta x_\mathrm{C} 
		+ \Delta x_\mathrm{B}
	\right) , \label{eq:s_PD_1B_2}\\
s_{\mathrm{PD_{2A}}} = & \frac{1}{1 + G_\mathrm{2A}}
	\left(
		- \Delta x_\mathrm{C} 
		+ \Delta x_\mathrm{A} 
	\right),   \label{eq:s_PD_2A_2} \\
s_{\mathrm{PD_{3B}}} = & \begin{cases}
    - \Delta x_\mathrm{A} + \Delta x_\mathrm{B} \ \ \ (|G_\mathrm{1B}| \ll 1, \ |G_\mathrm{2A}| \ll 1) \\ 
    \frac{1}{1+G_\mathrm{1B}}\left(-\Delta x_\mathrm{A} + \Delta x_\mathrm{B}\right) \ \ \ (G_\mathrm{2A} \simeq G_\mathrm{1B})
  \end{cases} \label{eq:s_PD_3B_2}
\end{align}
Here, we assume that all arm cavities have almost the same length, $L$. Equations (\ref{eq:s_PD_1B_2})-(\ref{eq:s_PD_3B_2}) indicate that differential signals between two cavities, which include gravitational wave signals, can be obtained in the DPDFPI.

Here, we explain how to make the Laser 3 resonate with the Cavity B. Let us consider the frequency offset of the Laser 2, $\Delta \nu_3$, from the resonant frequency of the Cavity B, $N\frac{c}{2L_\mathrm{B}}$ ($N\in\mathbb{N}$). $\Delta \nu_3$ is expressed by
\begin{equation}
\Delta \nu_3 \equiv \nu_3 -  N\frac{c}{2L_\mathrm{B}}, \label{eq:defDeltanu3}
\end{equation}
where $L_\mathrm{B}$ is the length of the Cavity B and $c$ is the speed of light. Since the frequency of the Laser 3 is controlled to follow the resonant frequency of the Cavity A, $\nu_3$ is written as
\begin{equation}
\nu_3 = N^\prime\frac{c}{2L_\mathrm{A}} \ \ \ (N^\prime\in\mathbb{N}),
\end{equation}
where $L_\mathrm{A}$ is the length of the Cavity A. Hence, $\Delta \nu_3$ is denoted as
\begin{equation}
\Delta \nu_3 = N^\prime\frac{c}{2L_\mathrm{A}} - N\frac{c}{2L_\mathrm{B}}.
\end{equation}
When the length of the Cavity A is $L_\mathrm{A} = L_\mathrm{B} + \Delta L$ ($|\Delta L| \ll L_\mathrm{B}$), $\Delta \nu_3$ is written as
\begin{equation}
\Delta \nu_3 = \frac{c}{2L_\mathrm{B}}\left(\Delta N - N\frac{\Delta L}{L_\mathrm{B}}\right), \label{eq:DelNu}
\end{equation}
where $\Delta N \equiv N^\prime - N$. By choosing $\Delta N$ to be the proper integer to $-N\frac{\Delta L}{L_\mathrm{B}}$ by adjusting $\nu_3$, $\Delta \nu_3$ is constrained to be within 
\begin{equation}
|\Delta \nu_3| \leq \frac{c|\Delta L|}{2L_\mathrm{B}^2}. \label{eq:AbsDelNu}
\end{equation}
This is because, when we change $\Delta N \rightarrow \Delta N +1$ and $N \rightarrow N +1$ , $\Delta \nu_3$ is changed by the difference of the free spectral range of the Cavity A and B as
\begin{equation}
\left| \frac{c}{2L_\mathrm{A}} - \frac{c}{2L_\mathrm{B}} \right| = \frac{c|\Delta L|}{2L_\mathrm{B}^2}. \label{eq:freqreso}
\end{equation}
Here, we use the fact of $|\Delta L| \ll L_\mathrm{B}$.

In order to resonate the Cavity B with the Laser 3, $\Delta \nu_3$ has to be well within the linewidth of the Cavity B. Consequently, eq. (\ref{eq:AbsDelNu}) indicates that we need to adjust $\Delta L$ for the resonance. For example, in DECIGO and B-DECIGO, the cavity linewidth is 15 Hz \cite{Kawamura2020}. Thus we need to adjust the cavity length to be $\Delta L \ll 100$ km for DECIGO ($L = 1000$ km) and $\Delta L \ll 1$ km for B-DECIGO ($L = 100$ km) for the operation of the DPDFPI. 
Moreover, for the reduction of the interferometer noise coupled with the detuning, e.g. laser intensity coupling noise, the requirement for $\Delta L$ can be strict depending on the sensitivity requirement. 
For example, if the laser intensity coupling noise is considered, $\Delta L< \frac{2L^2_\mathrm{B}h_\mathrm{req}}{cI_\mathrm{RIN}}\frac{}{}\nu_3$ where $h_\mathrm{req}$ is the sensitivity requirement, and $I_\mathrm{RIN}$ is a relative intensity noise of the input laser.

\section{Experimental setup for the demonstration of the DPDFPI}

\begin{figure}[htb]
  \centering
  \includegraphics[width=100mm]{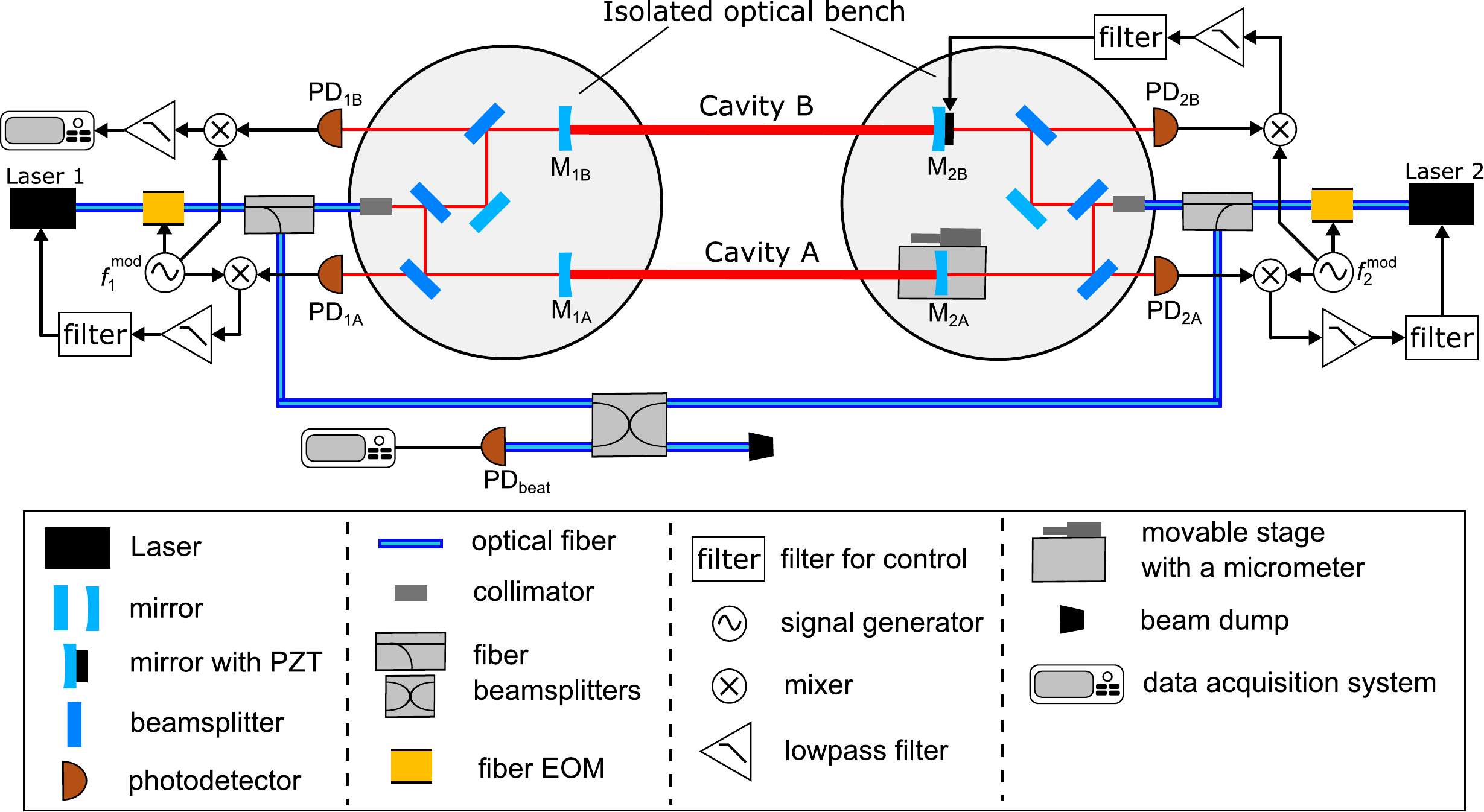}
  \caption{Schematic of the DPDFPI experiment.}
  \label{fig:setup}
\end{figure}

Since the DPDFPI is the new interferometer configuration, the experimental demonstration of the DPDFPI is necessary. Especially, the following two points should be confirmed: first, dependence of the laser frequency offset from the cavity resonant frequency on the cavity length difference discussed in the previous section, and, second, the operation of the DPDFPI, i.e. the measurement of the differential cavity displacement signal. 
For the experimental demonstration, we construct the prototype of the DPDFPI. 
Figure \ref{fig:setup} shows the schematic of the experiment of the DPDFPI prototype. In the DPDFPI prototype, only two cavities are used since the operation of the DPDFPI can be confirmed by evaluating the correlation of the signals measured with two lasers. Even in this setup, the key feature of the DPDFPI, i.e. the necessity of the cavity length adjustment for the interferometer operation, still remains. Thus, we need to adjust the length of the Cavity A against the length of the Cavity B to operate the interferometer.

In the experiment, we use two laser sources, Koheras AdjustiK C15 (Laser 1) and Koheras BASIK X15 (Laser 2), with a wavelength of 1550 nm. The output power of the Laser 1 and the Laser 2 are 10 mW and 30 mW, respectively. The laser beams are phase modulated with electro-optic modulators for the Pound--Drever--Hall technique \cite{Drever1983}. After the electro-optic modulators, the laser beams are splitted into two ways. One beam is injected to the main interferometer, i.e. the DPDFPI, and another beam is injected to the auxiliary interferometer for the cavity absolute length measurement, which is explained in \ref{Appdx:Length}. The length of the two cavities is measured to be $0.55340 \pm 0.00001$ m. The length of the Cavity A is able to be adjusted using the stage with the movable stage. For the main cavities, the lasers are injected from both sides. The main cavities are composed of the mirrors having the same specification. Their radius of curvature is 2 m and their amplitude reflectivity is 0.992. The reflected and transmitted beams from the cavities are measured with the photodetectors and the cavity longitudinal signals are obtained with the Pound--Drever--Hall technique \cite{Drever1983}. The two laser frequencies are shifted by one free spectral range of the Cavity A, $\frac{c}{2L_\mathrm{A}}$. The cavity mirrors are placed on the optical bench that is isolated with the rubber stack. The resonant frequency of the optical bench is about 10 Hz.

\section{Results and discussions of the demonstration experiment}

\begin{figure}[htb]
  \centering
  \includegraphics[width=100mm]{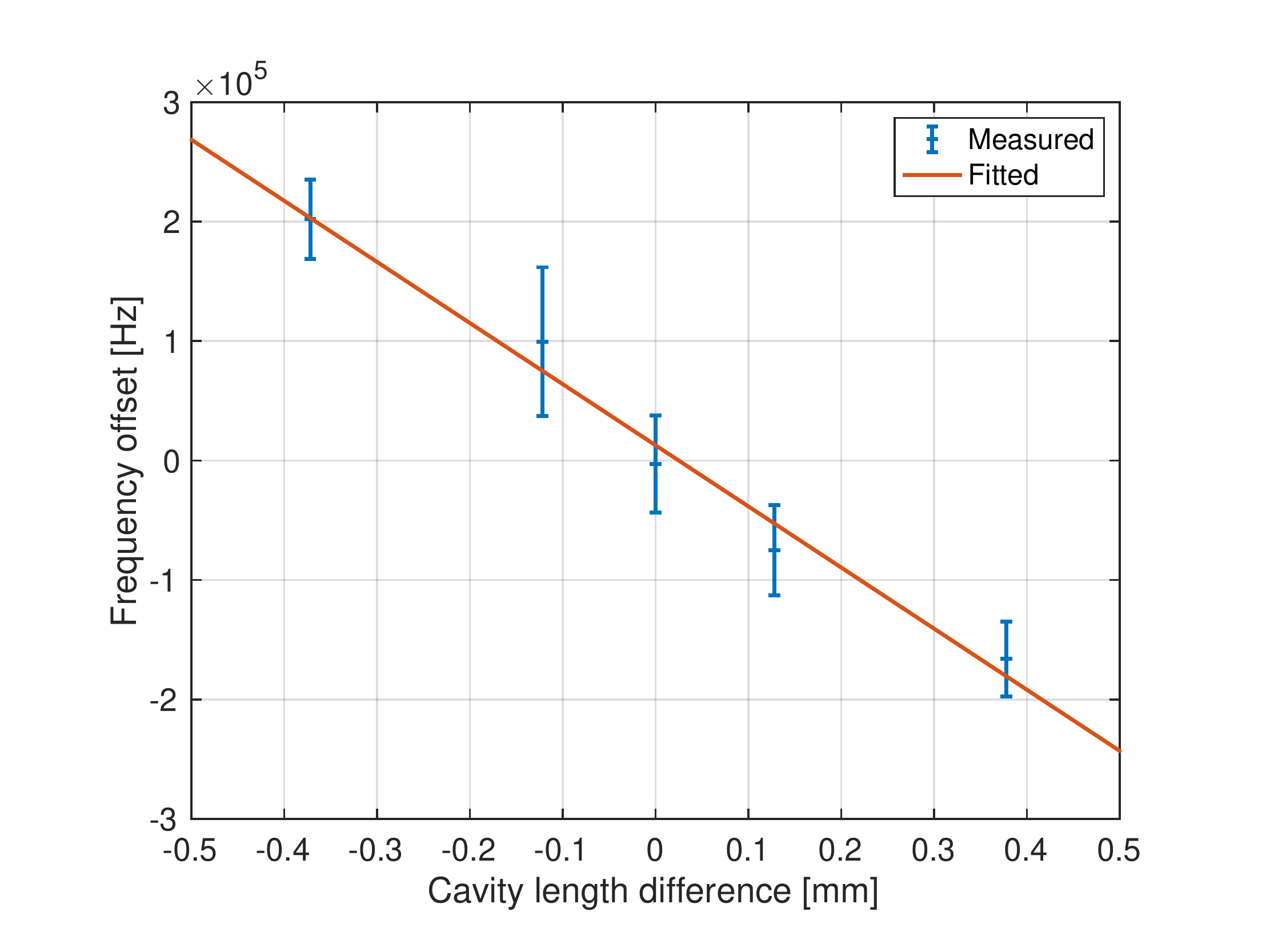}
  \caption{Measured frequency difference, $\Delta \nu$, between the frequency of the Laser 1 and the resonant frequency of the Cavity B. The solid line is determined by fitting the measured data.}
  \label{figC6:PDHoffset3}
\end{figure}

We investigated the dependence of the frequency difference, $\Delta \nu$, between the frequency of the Laser 1 and the resonant frequency of the Cavity B on the cavity length difference between the Cavity A and B. $\Delta \nu$ corresponds to $\Delta \nu_3$ in the previous discussion in Section \ref{ss:DPDFPI}. The measured result is shown in fig. \ref{figC6:PDHoffset3}. Figure \ref{figC6:PDHoffset3} shows that $\Delta \nu$ is shifted depending on the cavity length difference. From fig. \ref{figC6:PDHoffset3}, the relation between the frequency offset and the cavity length difference is determined to be $(-5.1 \pm 0.5)\times 10^8$ Hz/m by linear fitting the measured data. The determined value is consistent with the expected value, $(-4.8976 \pm 0.0002) \times10^8$ Hz/m, from eq. (\ref{eq:freqreso}) and the measured cavity length within the error ranges.



\begin{figure}[htb]
  \centering
  \includegraphics[width=100mm]{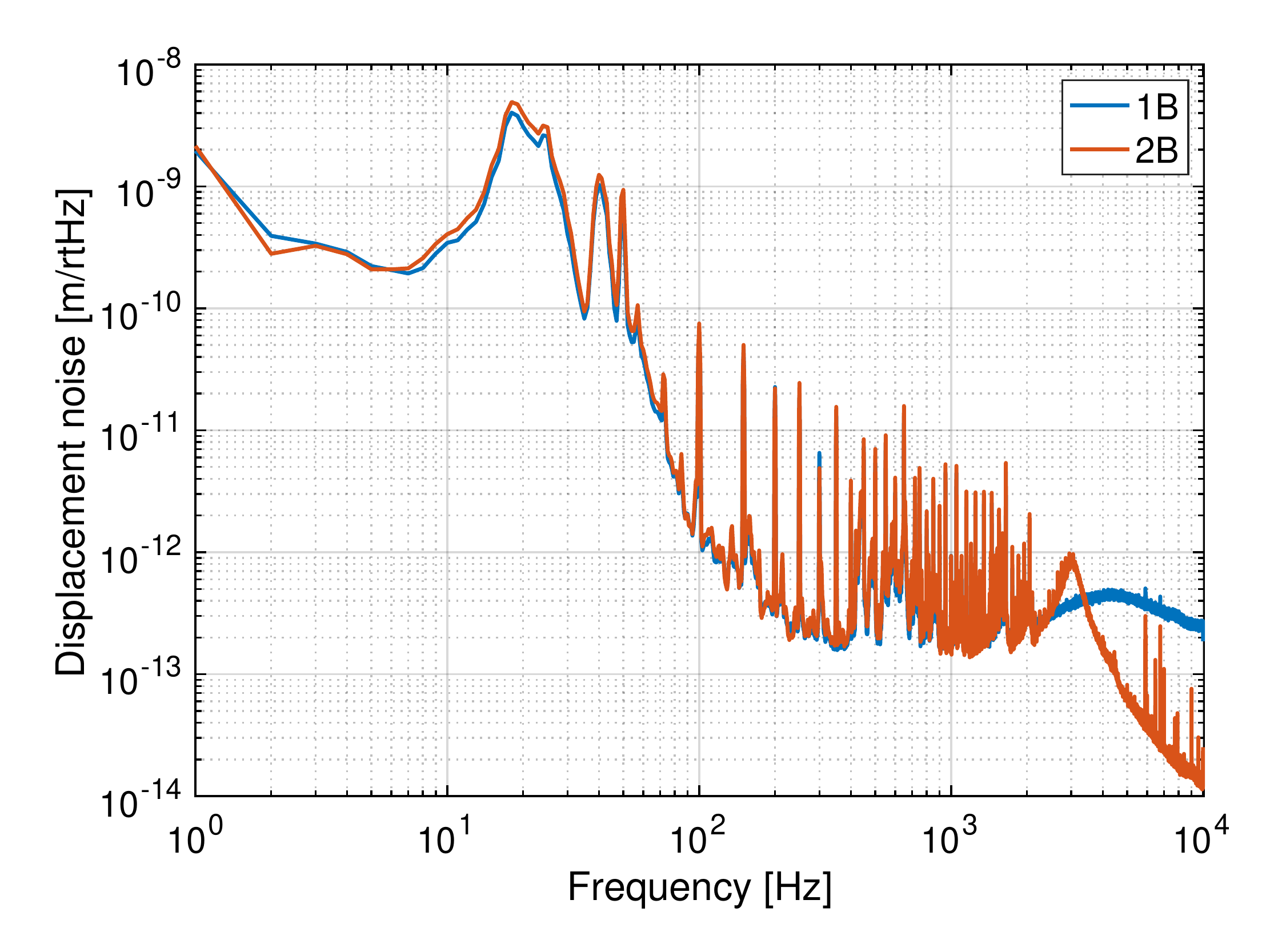}
  \caption{Calibrated noise spectra of the DPDFPI measured with the PD$_\mathrm{1B}$ and the PD$_\mathrm{2B}$.}
  \label{figC6:Noise}
\end{figure}

After the cavity length adjustment, the noise spectra of the interferometer were measured as shown in figure \ref{figC6:Noise}. The `1B' and `2B' curves represent the spectra measured with PD$_\mathrm{1B}$ and PD$_\mathrm{2B}$, respectively. In the proper condition of the open loop gain, the spectra shown in fig. \ref{figC6:Noise} indicate the differential length fluctuations of the Cavity A and the Cavity B as discussed in Section \ref{ss:DPDFPI}. Note that the unity gain frequency of the frequency control loop of the Laser 1 and the Laser 2 were measured to be 2.9 kHz and 4.0 kHz, respectively.

%

\begin{figure}[htb]
  \centering
  \includegraphics[width=100mm]{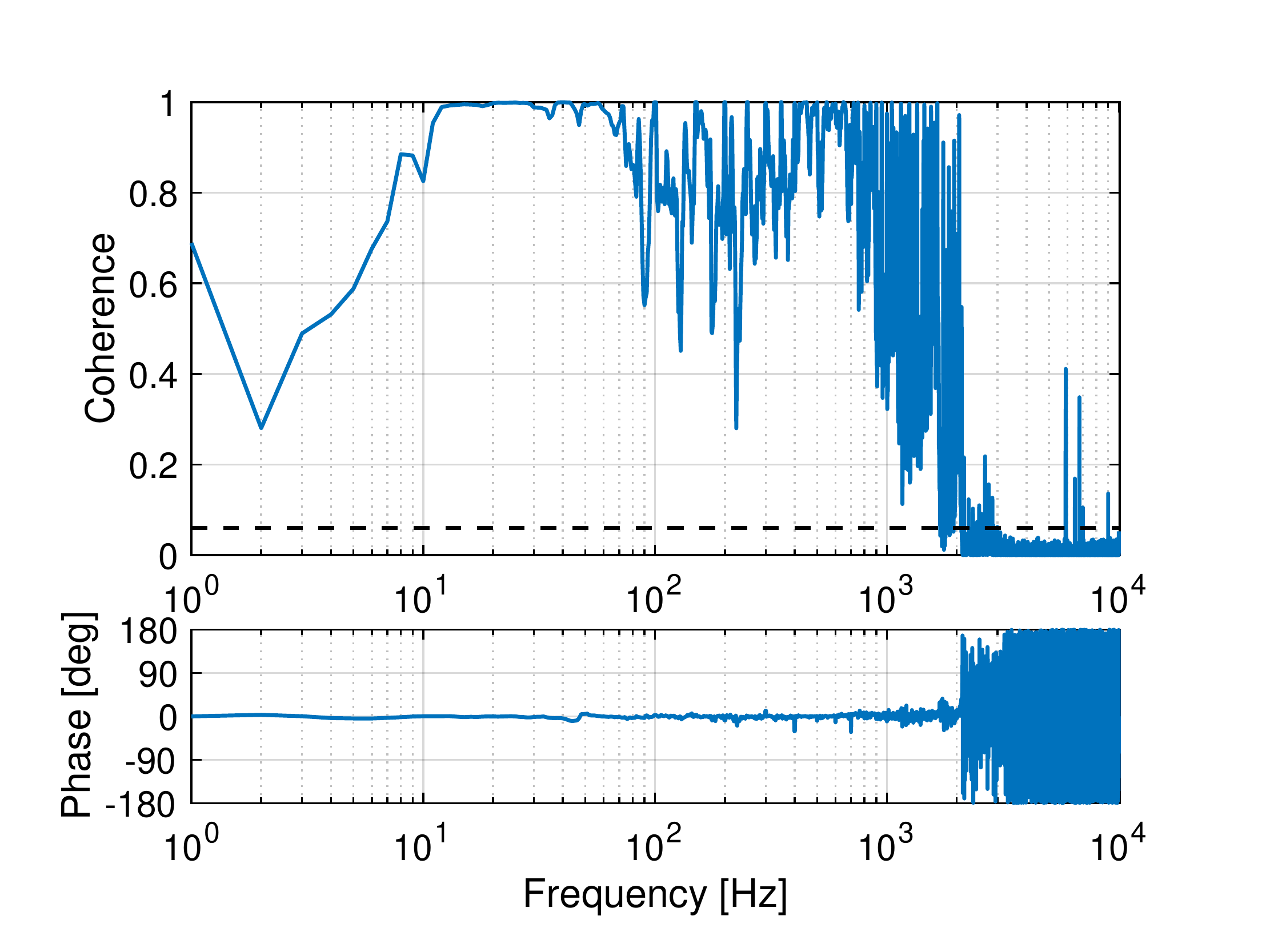}
  \caption{Measured magnitude-squared amplitude (upper panel) and phase (lower panel) of the coherence between the signals from the PD$_\mathrm{1B}$ and the PD$_\mathrm{2B}$. The dashed line is the significance threshold of the coherence, 0.06 \cite{Thompson1979}.}
  \label{figC6:coherence}
\end{figure}

If the DPDFPI in fig. \ref{fig:setup} is properly operated, the correlated differential signals between the Cavity A and the Cavity B are measured with the PD$_\mathrm{1B}$ and the PD$_\mathrm{2B}$. To check the correlation, the coherence between the signals from the PD$_\mathrm{1B}$ and the PD$_\mathrm{2B}$ is calculated as shown in fig. \ref{figC6:coherence}. Given the average number of 100 for the calculation, the 95\% significance threshold of the coherence is 0.06 \cite{Thompson1979}. Thus, fig. \ref{figC6:coherence} indicates that the two signals are coherent below $\sim$1.5 kHz as expected from the unity gain frequency of the frequency control loops of the Laser 1 and the Laser 2. Hence, the DPDFPI is conceived properly operated.


\section{Conclusion}

In this paper, we presented the working principle of the DPDFPI for the first time. For the operation of the DPDFPI, the absolute length adjustment is necessary. Moreover, using the 55-cm-long DPDFPI prototype, we demonstrated the operation of the DPDFPI and confirmed that adjustment of the absolute arm length reduced the cavity detuning as expected with our formulation. This work provides the proof of concept of the DPDFPI for application to the future Fabry--Perot type space gravitational wave antennas. 


\section*{Acknowledgements}

We thank Kiwamu Izumi, Tomofumi Shimoda, Yutaro Enomoto, and Ayaka Shoda for fruitful discussions. We appreciate the technical support from Shigemi Otsuka, and Togo Shimozawa in the machine shop of Graduate School of Science, The University of Tokyo. We also thank Kunihiko Hasegawa and Takaharu Shishido for preparing some mechanical components in the experiment. This work is financially supported by JSPS Grant-in-Aid for Scientific Research (A) No. 15H02087 and JSPS KAKENHI Grant No. JP17J01176.

\appendix

\section{Cavity absolute length measurement} \label{Appdx:Length}

Here, we explain how to measure the absolute length of the cavity in our experiment. 
We adopt a similar scheme to \cite{Araya1999a, Stochino2012a}. In the dual-pass cavity, two different lasers resonate with a cavity. Thus, by measuring the frequency difference between the two lasers, we are able to determine the cavity free spectral range, which is related with the cavity length. For example, in our experimental setup shown in fig. \ref{fig:setup}, the frequency difference of the Laser 1 and the Laser 2, $\nu_\mathrm{diff}$, is expressed as
\begin{equation}
\nu_\mathrm{diff} = n\frac{c}{2L_\mathrm{A}} = n^\prime\frac{c}{2L_\mathrm{B}} \ \ (n, n^\prime\in\mathbb{N}).
\end{equation}
Note that $\frac{c}{2L_\mathrm{A}}$ and $\frac{c}{2L_\mathrm{B}}$ are the free spectral range of the Cavity A and the Cavity B, respectively. When we know $n$ ($n^\prime$), the cavity length $L_\mathrm{A}$ ($L_\mathrm{B}$) can be determined by measuring $\nu_\mathrm{diff}$. $\nu_\mathrm{diff}$ is measured by observing the interference between the two lasers. In this work, $n$ and $n^\prime$ are set to be 1 and the interference signal between the Laser 1 and the Laser 2 is measured with the PD$_\mathrm{beat}$ in fig. \ref{fig:setup}.

It is worth noting that, even in the setup shown in fig. \ref{fig:DPDFPI}, the cavity length can be measured with almost the same scheme used in fig. \ref{fig:setup}. One difference is using the injected and transmitted laser of the cavity to obtain the interference signal. Although two injected lasers are interfered in this work, it is challenging to do the same thing in the space detectors where two laser sources are placed in the distant satellites. 

\section*{References}

\bibliography{library}

\end{document}